\PassOptionsToPackage{dvipsnames, usenames}{xcolor}

\documentclass[fleqn,usenatbib]{mnras_tex_edited}

\usepackage{amssymb,amsmath,verbatim,mathtools,needspace,etoolbox,graphicx,physics,microtype,natbib,url,hyperref,mathrsfs,bm}
\usepackage{orcidlink}

\def\aj{{Astron.~J.}}

\def\apj{{Astrophys.~J.}}\def\apjl{{Astrophys.~J.~Lett.}}
\def\apjs{{Astrophys.~J.~Supp.}}

\def\aap{{Astron.~Astrophys.}}

\def\mnras{{Mon.~Not.~R.~Astron.~Soc.}}

\def\prd{{Phys.~Rev.~D}}

\def\prx{{Phys.~Rev.~X}}
\def\prl{{Phys.~Rev.~Lett.}}
\def\prr{{Phys.~Rev.~Res.}}
\def\pasa{{Publ. Astron. Soc. Aust.}}

\def\nat{{Nature}}

\def\lrr{{Living Rev. Relativ.}}
\def\cqg{{Classical~Quant.~Grav.}}
\def\natastro{{Nat. Astron.}}

\definecolor{linkcolor}{rgb}{0.0,0.3,0.5}
\hypersetup{colorlinks=true, linkcolor=linkcolor, citecolor=linkcolor, filecolor=linkcolor, urlcolor=linkcolor}

\renewcommand{\vec}[1]{\boldsymbol{#1}}

\newcommand{\bham}{School of Physics and Astronomy \&	 Institute for Gravitational Wave Astronomy, University of Birmingham, Birmingham,\vspace{-0.05cm}\\$\;$B15 2TT, UK}
\newcommand{\milan}{Dipartimento di Fisica `G. Occhialini', Universit\'a degli Studi di Milano-Bicocca, Piazza della Scienza 3, 20126 Milano, Italy}
\newcommand{\infn}{INFN, Sezione di Milano-Bicocca, Piazza della Scienza 3, 20126 Milano, Italy}
\newcommand{\padova}{Physics and Astronomy Department Galileo Galilei, University of Padova, Vicolo dell'Osservatorio 3, I--35122 Padova, Italy}
\newcommand{\infnpadova}{INFN--Padova, Via Marzolo 8, I--35131 Padova, Italy}
\newcommand{\heidelberg}{Universit\"at Heidelberg, Zentrum f\"ur Astronomie, Institut f\"ur Theoretische Astrophysik, Albert-Ueberle-Str. 2, D--69120 Heidelberg,\vspace{-0.05cm}\\$\;$Germany}

\title[One to many]{One to many: comparing single gravitational-wave events to astrophysical populations}

\author[M. Mould et al.]{
Matthew Mould\,\orcidlink{0000-0001-5460-2910}$^{1}$\thanks{\href{mailto:mmould@star.sr.bham.ac.uk}{mmould@star.sr.bham.ac.uk}},
Davide Gerosa\,\orcidlink{0000-0002-0933-3579}$^{2,3,1}$,
Marco Dall'Amico\,\orcidlink{0000-0003-0757-8334}$^{4,5}$,
Michela Mapelli\,\orcidlink{0000-0001-8799-2548}$^{4,5,6}$
\medskip
\\
$^{1}$\bham\\
$^{2}$\milan\\
$^{3}$\infn\\
$^{4}$\padova\\
$^{5}$\infnpadova\\
$^{6}$\heidelberg
}

\pubyear{\the\year{}}

\allowdisplaybreaks

\begin{document}

\label{firstpage}
\pagerange{\pageref{firstpage}--\pageref{lastpage}}
\maketitle

\begin{abstract}
Gravitational-wave observations have revealed sources whose unusual properties challenge our understanding of compact-binary formation.
Inferring the formation processes that are best able to reproduce such events may therefore yield key astrophysical insights.
A common approach is to count the fraction of synthetic events from a simulated population that are consistent with some real event.
Though appealing owing to its simplicity, this approach is flawed because it neglects the full posterior information, depends on an ad-hoc region that defines consistency,
and fails for high signal-to-noise detections.
We point out that a
statistically consistent solution is to compute the posterior odds between two simulated populations, which crucially is a relative measure, and show how to include the effect of observational biases by conditioning on source detectability.
Applying the approach to several gravitational-wave events and simulated populations, we assess the degree to which we can conclude model preference not just between distinct formation pathways but also between subpopulations within a given pathway.
\end{abstract}


\section{Introduction}
\label{sec: Introduction}

In the first three observing runs~\citep{2019PhRvX...9c1040A, 2021PhRvX..11b1053A, 2021arXiv210801045T, 2021arXiv211103606T} of the LIGO~\citep{2015CQGra..32g4001L}, Virgo~\citep{2015CQGra..32b4001A}, and KAGRA~\citep{2019NatAs...3...35K} gravitational-wave (GW) interferometers there have been nearly 100 detection candidates of transient signals originating from stellar-mass compact-object binaries.
Despite the growing catalogue of observations, there are still new events whose unusual or unique properties make them especially interesting.
For example:
GW200129 is a binary black hole (BH) merger whose remnant is constrained to have a large gravitational kick~\citep{2022PhRvL.128s1102V} and which may (or may not, \citealt{2022PhRvD.106j4017P}) exhibit evidence for relativistic spin precession (\citealt{2022Natur.610..652H});
GW190521 consists of two BHs with particularly large component masses~\citep{2020PhRvL.125j1102A} and features possible signatures of orbital eccentricity~\citep{2020ApJ...903L...5R, 2022NatAs...6..344G};
GW190412 confidently contains two stellar-mass BHs but with clearly unequal masses~\citep{2020PhRvD.102d3015A};
and for GW190814~\citep{2020ApJ...896L..44A} the lighter component is either an unusually heavy neutron star or light BH, making its formation with very unequal masses difficult to explain.

The atypical properties of these events have important implications for their progenitors, as astrophysical formation environments must facilitate the production of such sources.
In the case of GW190521, for example, several possibilities have been proposed to explain its formation~\citep{2020ApJ...900L..13A, 2020ApJ...903L...5R, 2020ApJ...904L..26F, 2021ApJ...907L...9N, 2021PhRvL.126h1101B, 2021PhRvD.104h3036O, 2022ApJ...924...79E, 2022NatAs...6..344G, 2023NatAs...7...11G}.
This event is particularly problematic because, while it confidently has a large total mass, the short signal duration leads to ambiguity in the nature of the source~\citep{2021PhRvD.104j3018B, 2023MNRAS.519.5352R}.

Given the vast array of possible -- if not plausible -- formation scenarios, a suitable metric is needed to asses the relative degree of consistency between GW measurements and different models of formation.
A common method of building such models is to simulate a population of mergers.
One may therefore be interested in investigating whether or not such simulations are capable of reproducing populations that are consistent with a given GW event, and conversely whether a given event is consistent with a synthetic population
(see, e.g.,
\citealt{
2020PhRvL.125j1103G,
2021MNRAS.505..339M,
2021MNRAS.508.3045D,
2021ApJ...908..194T,
2022ApJ...933...86Z,
2022ApJ...938...45B,
2023ApJ...954...23Z}).

But how does one consistently compare a \textit{distribution} of simulated sources with a \textit{single} observed event?

A typical approach proceeds by
(i) constructing a synthetic population of mergers (e.g., through population synthesis simulations),
(ii) defining a region in the parameter posterior of a GW event of interest [e.g., 90\% credible intervals (CIs) in masses and spins], and
(iii) counting the number of simulated mergers inside this region.
A high fraction of simulated samples lying inside the chosen region is taken as an indication that the detected GW event is consistent with the simulated formation scenario.
Conversely, one could define the confidence region in terms of the simulated mergers and count the number of posterior samples from the event inside that region.

However, this method is at best approximate because:
for simplicity, the confidence region is typically assembled heuristically as the product of marginalized one-dimensional (1D) CIs, and the result depends on the choice of quantile (e.g., 90\%);
density gradients inside this region are neglected;
posterior support outside the region is neglected;
the influence of the original parameter estimation (PE) prior is not addressed;
and there is an ambiguity in which of the two comparisons to make, as they do not agree.

These issues become especially clear in the limit of high signal-to-noise ratios (SNRs).
For a high-SNR detection the posterior distribution is narrow and, therefore, the fraction of simulated mergers inside any region of posterior support will be arbitrarily close to zero, even if the posterior lies entirely within the extremal simulated sources.
Conversely, since the posterior support is bounded well within the range of simulated sources, the fraction of posterior samples inside any given region of the simulated population will be arbitrarily close to unity.

In this work we point out a statistically consistent alternative provided by working entirely within the framework of Bayesian PE and model comparison.
In short, the posterior odds can be used when comparing the posterior of a GW event with a model of formation, the latter acting as a Bayesian prior.
An important point is that one should not ask to what extent a GW event matches a particular model of formation in absolute terms, and should instead quantify how much more or less likely one model is over another.
The statistics required to perform Bayesian PE and model comparison (and hierarchical inference) have already been extensively reviewed elsewhere (see, e.g., \citealt{2019PASA...36...10T, 2019MNRAS.486.1086M, 2022hgwa.bookE..45V}).
We summarize the salient details and point out that the analysis required is simple and, in the context of GW astronomy, readily performed with publicly available data products combined with a predictive astrophysical model.

\subsection{Practical summary}
\label{sec: Practical summary}

One first identifies a GW event of interest, perhaps because it is an outlier with respect to other observations
\citep{
2020ApJ...891L..31F, 2022ApJ...926...34E, 2022PASA...39...25R}.
Then, irrespective of the remaining catalogue, we wish to compare models of formation for this individual event.

Performing a comparison between astrophysical population models -- say, $A$ and $B$ -- and deciding which is more consistent with the observed GW event with parameters $\vec{\theta}$ (e.g., masses, spins, redshift, etc.) requires computing the likelihoods that the GW data are drawn from those populations. 
The required steps are:
\begin{enumerate}
\item
Simulate an astrophysical set of mergers from model $A$; $\{\vec{\theta}_{Aj}\}_{j=1}^{N_A}$.
\item
Construct a probability density estimate of the simulated merger source properties; $\pi(\vec{\theta}|A)$.
\item
Evaluate whether each simulated source is detectable or not; $\{P(\mathrm{det}|\vec{\theta}_{Aj})\}_{j=1}^{N_A}$.
\item
Repeat steps (i)--(iii) for model $B$.
\item
Download the PE samples for the GW event of interest; $\{\vec{\theta}_i\}_{i=1}^N$.
If not already known, construct a density estimate for the uninformative PE prior; $\pi(\vec{\theta}|U)$.
\item
Compute the detectability-weighted Bayes factor
\begin{align}
\mathcal{D}_{A/B}
=
\frac{ P(\mathrm{det}|B) } { P(\mathrm{det}|A) }
\mathcal{B}_{A/B}
\, ,
\label{eq: Bayes factor det intro}
\end{align}
\end{enumerate}
where, by summing over simulated sources, the average detection probability for population $A$ is
\begin{align}
P(\mathrm{det}|A)
=
\frac{ 1 } { N_A } \sum_{j=1}^{N_A} P(\mathrm{det}|\vec{\theta}_{Aj})
\label{eq: pdet intro}
\end{align}
and, by summing over posterior samples for the real event, the Bayes factor is
\begin{align}
\mathcal{B}_{A/B}
=
\frac{ \sum_{i=1}^N \pi(\vec{\theta}_i|A) / \pi(\vec{\theta}_i|U) } { \sum_{i=1}^N \pi(\vec{\theta}_i|B) / \pi(\vec{\theta}_i|U) }
\, .
\label{eq: Bayes factor intro}
\end{align}
If $\mathcal{D}_{A/B} > 1$ ($<1$) then, assuming equal prior odds, the detected GW event is a posteriori more consistent with population $A$ ($B$) -- the key statement of interest.
Since this posterior odds (or Bayes factor) is inherently a relative measure, considering a single model is not possible; the absence of a second astrophysical model implies a comparison to the original PE prior.

In Sec.~\ref{sec: Comparing a detected event to astrophysical populations}, we formally define the heuristic fraction used previously and demonstrate why it is unsuitable.
We summarize Bayesian model comparison in the present context of GW astronomy and derive the above quantities step by step.
In Sec.~\ref{sec: Examples in gravitational-wave astronomy}, we apply this formalism to several stellar-mass mergers detected by LIGO/Virgo and assess to what degree we can conclude that they are consistent with various models of formation.
We end on some final thoughts in Sec.~\ref{sec: Conclusions}.
Code to reproduce these results is publicly available (\href{https://github.com/mdmould/popodds}{github.com/mdmould/popodds}; \citealt{matthew_mould_2023_7647704}).

\section{Bayesian model comparison}
\label{sec: Comparing a detected event to astrophysical populations}

Using Bayes' theorem, the measured properties of a GW event are given by the posterior distribution
\begin{align}
p(\vec{\theta}|\vec{d},U)
=
\frac{ \mathcal{L}(\vec{d}|\vec{\theta}) \pi(\vec{\theta}|U) } { \mathcal{Z}(\vec{d}|U) }
\, ,
\label{eq: posterior}
\end{align}
where
$\mathcal{L}(\vec{d}|\vec{\theta})$
is the likelihood that the vector of waveform parameters $\vec{\theta}$ (e.g., masses, spins, distance, sky position, etc.) produce the GW data $\vec{d}$, $\pi(\vec{\theta}|U)$ represents our typically uninformative prior belief on what the waveform parameters could be, and
\begin{align}
\mathcal{Z}(\vec{d}|U)
=
\int \mathcal{L}(\vec{d}|\vec{\theta}') \pi(\vec{\theta}'|U) \dd\vec{\theta}'
\label{eq: evidence}
\end{align}
is the marginal likelihood -- or evidence -- that normalizes the posterior by ensuring
$\int p(\vec{\theta}|\vec{d},U) \dd\vec{\theta} \equiv 1$
and represents the probability that the chosen PE prior produces the observed data.

We only explicitly condition prior-dependent terms on the uninformative model $U$ to represent their dependence on the shape of the chosen prior distribution. The result also depends on, e.g., the waveform model that in turn determines those parameters $\vec{\theta}$ under consideration. We omit these extra model conditionals that we consider fixed to avoid unnecessarily lengthy expressions.

Given such inference on the binary properties $\vec{\theta}$, the question is whether the measurement $p(\vec{\theta}|\vec{d},U)$ is consistent with a particular model of binary formation. We describe such approaches in the following sections.

\subsection{Heuristic approximation}
\label{sec: Heuristic approximation}

We begin with the heuristic fraction that is often used to assess the consistency of GW events with populations models.
It is computed by counting the number of sources simulated from some astrophysical model that lie inside a chosen region $R$ of the measured properties $p(\vec{\theta}|\vec{d},U)$.
Such simulations can be phrased as a finite set of draws from an unknown underlying astrophysical distribution $\pi(\vec{\theta}|A)$ that describes the result of the simulation.
The fraction inside $R$ is
\begin{align}
f_A(R)
=
\frac{ N_A(R) } { N_A }
\approx
\int_R \pi(\vec{\theta}|A) \dd\vec{\theta}
=
\int
\mathcal{I}(\vec{\theta},R) \pi(\vec{\theta}|A) \dd\vec{\theta}
\, ,
\label{eq: heuristic fraction}
\end{align}
where $N_A$ is the total number of simulated sources, $N_A(R)$ is the number of those sources that fall within the targeted region $R$ of the posterior, and $\mathcal{I}(\vec{\theta},R)$ is an indicator function equal to unity if $\vec{\theta} \in R$ and zero otherwise.
Large values of $f_A(R)$ can be taken as an indication that the event posterior is consistent with the simulated model, though the value at which one may define significance is arbitrary.
A similar conclusion holds conversely for the fraction of GW posterior samples lying inside a parameter region defined in terms of the simulated sources;
this fraction is not usually used as population models are typically broader than GW event posteriors, meaning the fraction will always be close to unity.

The region $R$ is often constructed heuristically by joining the 90\% CIs of each ID marginalization of the posterior.
However, this choice is complicated by the addition of higher dimensions and is not necessarily a good approximation of the true full-dimensional 90\% credible region.
Additionally, as one includes more parameters in the comparison while keeping the simulation size fixed, the curse of dimensionality means that fewer simulated sources will fall within the chosen region, leading to the naive conclusion that the model is harder to reconcile with the data.

\begin{figure}
\includegraphics[width=\columnwidth]{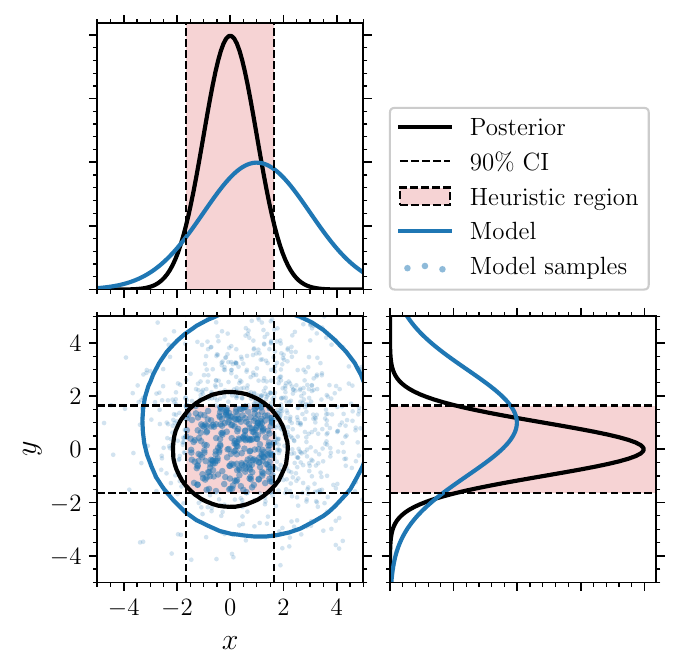}
\caption{
Illustration of the heuristic fraction.
The posterior from a single observation
(solid black) and simulated fiducial model (solid blue) are 2D normal distributions.
In the lower left panel are their 90\% credible regions while the diagonal panels display their 1D marginals.
The dashed lines are located at the boundaries of the symmetric 90\% CIs of the marginalized posteriors, while the red region highlights the heuristic region assembled as the product of the individual intervals.
Blue points represent sources simulated from the model, with those inside the red region being counted for the heuristic fraction.
}
\label{fig: cartoon}
\end{figure}

We illustrate the setup of this heuristic fraction in Fig.~\ref{fig: cartoon}.
We take different two-dimensional (2D) normal distributions in $(x,y)$ as both the posterior and simulated population model.
In red we highlight the heuristic region $R$ assembled as the product of the 1D 90\% posterior CIs; note that this region does not match the joint-dimensional 90\% credible region of the posterior.
Sources simulated from the fiducial model are plotted as blue markers, with those within the heuristic region highlighted in bold.
The heuristic fraction is found as the fraction of the simulated points that lie inside the red square.
This computation neglects any posterior support outside the red region and also neglects variation of the posterior density inside it.
One can see that if the posterior were to become more precise (due to, e.g., a higher SNR), the red region would shrink and fewer simulated points lie inside it, even though the posterior is fully enclosed within the simulated model.
Additionally, the heuristic fraction is an absolute measure for a single model in that the influence of the original prior on the posterior is not accounted for.
As such, it is not clear when a value of $f_A(R)$ should be considered significant.

\subsection{Bayes factors}
\label{sec: Bayes factors}

We now seek a statistically consistent alternative to the heuristic fraction.
Namely, one can perform Bayesian model selection to find which models are preferred relative to others.

Starting from the GW event posterior in Eq.~(\ref{eq: posterior}), we can reconstruct the posterior that would have been inferred had we imposed our informative astrophysical model $A$ as a prior by re-weighting the posterior that we do have.
Since the likelihood is unchanged, the new posterior may be written as
\begin{align}
p(\vec{\theta}|\vec{d},A)
=
p(\vec{\theta}|\vec{d},U)
\frac{ \pi(\vec{\theta}|A) } { \pi(\vec{\theta}|U) }
\frac { \mathcal{Z}(\vec{d}|U) } { \mathcal{Z}(\vec{d}|A) }
\, .
\label{eq: reweighted posterior}
\end{align}
Given samples from the original posterior
$p(\vec{\theta}|\vec{d},U)$,
the new posterior is constructed using those samples with weights proportional to the prior ratio
$\pi(\vec{\theta}|A) / \pi(\vec{\theta}|U)$.

Since the posterior is normalized, integrating over $\vec{\theta}$ and rearranging yields the ratio of evidences -- also called the Bayes factor -- that quantifies the relative likelihood between the two prior models:
\begin{align}
\mathcal{B}_{A/U}
=
\frac
{ \mathcal{Z}(\vec{d}|A) }
{ \mathcal{Z}(\vec{d}|U) }
=
\int p(\vec{\theta}|\vec{d},U) \frac{ \pi(\vec{\theta}|A) } { \pi(\vec{\theta}|U) } \dd\vec{\theta}
\, .
\label{eq: Bayes factor}
\end{align}
A value of
$\mathcal{B}_{A/U} > 1$ ($<1$)
implies that the astrophysical model is more (less) likely to produced the GW data $\vec{d}$ than the original PE prior, with values further from unity indicating a stronger preference; a typical scale indicating a decisive conclusion is
$|\ln\mathcal{B}_{A/U}| \gtrsim 5$
\citep{1939thpr.book.....J}.
One may notice that this Bayes factor is proportional to the per-event term entering the usual GW population likelihood
\citep{
2019PASA...36...10T,
2019MNRAS.486.1086M,
2022hgwa.bookE..45V},
though here we consider population models to be fixed.
By reusing the posterior inferred under the default uninformative prior in this construction, we do not need to re-evaluate the computationally expensive likelihood
$\mathcal{L}(\vec{d}|\vec{\theta})$
or perform new PE runs.

We describe how to compute the Bayes factor in practice in Appendix~\ref{sec: Practical computation} and how to handle changes of variables in Appendix~\ref{sec: Missing and transformed parameters}.

Unlike the heuristic fraction, the Bayes factor is an explicit comparative measure between two models -- the original uninformative prior, $U$, and the new informative one, $A$.
By comparison with Eq.~(\ref{eq: Bayes factor}), we see that the heuristic fraction $f_A(R)$ in Eq.~(\ref{eq: heuristic fraction}) has the form of a Bayes factor with
$p(\vec{\theta}|\vec{d},U) / \pi(\vec{\theta}|U)$
replaced with the indicator function
$\mathcal{I}(\vec{\theta},R)$.
To be valid, this would require the posterior
$p(\vec{\theta}|\vec{d},U)$
to have no support outside the region $R$ and be equal to the original prior
$\pi(\vec{\theta}|U)$
inside $R$.
One case this holds is when the posterior only recovers the prior and the region $R$ is selected as the entire prior volume -- but this is clearly not true in general.
Equation~(\ref{eq: heuristic fraction}) also has the form of an evidence as in Eq.~(\ref{eq: evidence}) with the likelihood
$\mathcal{L}(\vec{d}|\vec{\theta})$
replaced with the indicator function (also an invalid approximation).
One could therefore consider the relative fraction
\begin{align}
\mathcal{F}_{A/U}(R)
=
\frac{ f_A(R) } { f_U(R) }
=
\frac{ \int \mathcal{I}(\vec{\theta},R) \pi(\vec{\theta}|A) \dd\vec{\theta} } { \int \mathcal{I}(\vec{\theta},R) \pi(\vec{\theta}|U) \dd\vec{\theta} }
\, ,
\label{eq: relative fraction}
\end{align}
as this is a comparative measure that has a similar interpretation to the Bayes factor with a natural scale: $\mathcal{F}_{A/U}(R) > 1$ ($< 1$) implies $A$ is favoured (disfavoured) with respect to $U$ over the heuristically chosen region $R$.
However, this relative heuristic fraction still suffers from the other issues enumerated in Sec.~\ref{sec: Heuristic approximation}.
We consider a simple example and compare the heuristic fraction, the relative fraction, and the Bayes factor in Appendix~\ref{sec: Simple example}.

\subsection{Posterior odds}
\label{sec: Posterior odds}

A comparison can be made between two different informative models -- say, $A$ and $B$ -- from neither of which we have inferred a posterior. The Bayes factor between $A$ and $B$ is just the ratio of the two Bayes factors with respect to the original uninformative prior $U$:
\begin{align}
\mathcal{B}_{A/B}
=
\frac{ \mathcal{Z}(\vec{d}|A) } { \mathcal{Z}(\vec{d}|B) }
=
\frac{ \mathcal{B}_{A/U} } { \mathcal{B}_{B/U} }
\, .
\label{eq: Bayes factor ratio}
\end{align}
Note, then, that while the Bayes factor between $A$ and $B$ does not depend on the original prior, its computation requires us to explicitly divide it out to compute $\mathcal{B}_{A/U}$ and $\mathcal{B}_{B/U}$ using Eq.~(\ref{eq: Bayes factor}) because it is the only one for which we have inferred a posterior.
Equation~(\ref{eq: Bayes factor ratio}) has the advantage of being independent of the normalization of the original PE prior, however, which only enters through $\mathcal{Z}(\vec{d}|U)$.
This fact is useful when only the shape and not the normalization of $\pi(\vec{\theta}|U)$ is known.
For example, in the context of GW astronomy one may choose uniform priors in detector-frame component masses but place limits (which may vary between events and waveform models) on the chirp mass and mass ratio
\citep{2023ApJS..267...29A},
resulting in a prior whose shape within the cut remains flat but whose overall normalization is altered.
This is irrelevant for inference but crucial for comparisons.

Finally, we arrive at the metric we sought: when comparing Bayesian models one may employ the posterior odds,
\begin{align}
\mathcal{O}_{A/B}
=
\frac{ p(A|\vec{d}) } { p(B|\vec{d}) }
=
\frac{ \pi(A) } { \pi(B) } \mathcal{B}_{A/B}
\, ,
\label{eq: posterior odds}
\end{align}
where $\pi(A)$ represents our prior on the hypothesis that $A$ is the correct model generating GW events and the ratio $\pi(A)/\pi(B)$ is the prior odds, i.e., how much weight we a priori assign to model $A$ over $B$.
In many situations we are unable to reasonably estimate such priors and instead default to equal prior odds, in which case $\mathcal{O}_{A/B} = \mathcal{B}_{A/B}$ and the posterior odds and Bayes factor coincide.
In this case, comparing two astrophysical models is typically more useful as the PE prior may often be preferred owing to its broad shape, even though it is not a realistic model of formation; this is where the prior odds would downweight preference for the PE prior if it could be reasonably chosen.

However, we stress that posterior measurements -- in particular, the posterior odds -- are the Bayesian metrics that should be used in these circumstances.
By defaulting to the Bayes factor, we must be mindful of the implicit assumption of equal prior odds and that an alternative conclusion may be drawn if we impose different model priors.

\subsection{Astrophysical populations and selection effects}
\label{sec: Astrophysical populations and selection effects} 

So far we have not considered selection effects -- the fact that detected GW events are subject to observational biases.
This means that certain signals are more easily detected than others, e.g., from sources which are closer to the detectors.
When considering an observed GW event, we know that this source must belong to the observable population -- not just the intrinsic astrophysical population -- and this knowledge can be folded into the model comparison.

We seek a generalization to Eq.~(\ref{eq: posterior odds}) for the posterior odds $\mathcal{O}_{A/B}$  that takes into account selection effects.
As in Eqs.~(\ref{eq: Bayes factor}) and (\ref{eq: Bayes factor ratio}), the central quantity we require is the evidence for an astrophysical model $A$, but now conditioned on detectability (det).
Using Bayes' theorem, we may write this as
\begin{align}
\mathcal{Z}(\vec{d}|A,\mathrm{det})
=
\frac
{ P(\mathrm{det}|\vec{d}) \mathcal{Z}(\vec{d}|A) }
{ P(\mathrm{det}|A) }
\, ;
\label{eq: evidence det}
\end{align}
cf. Eq.~(10) of \cite{2004AIPC..735..195L}.
The probability of successfully detecting a signal in the data is
$P(\mathrm{det}|\vec{d}) = 1$
for the detected GW event under consideration, where we are making the assumption of a deterministic detection threshold as a function of the data.
Given only the set of parameters $\vec{\theta}$, the source detection probability is found by marginalizing over the data distribution, i.e.,
\begin{align}
P(\mathrm{det}|\vec{\theta})
=
\int
P(\mathrm{det}|\vec{d}')
\mathcal{L}(\vec{d}'|\vec{\theta})
\dd\vec{d}'
\, .
\label{eq: Pdet source}
\end{align}
The prior-averaged detection probability
\begin{align}
P(\mathrm{det}|A)
=
\int
P(\mathrm{det}|\vec{\theta})
\pi(\vec{\theta}|A)
\dd\vec{\theta}
\label{eq: Pdet population}
\end{align}
is the fraction of sources from the population $A$ that is detectable.

In Eq.~(\ref{eq: evidence det}), the numerator contains information about the \emph{detected} source, being that its terms depend on the data, while the denominator accounts for the distribution of \emph{detectable} sources; all detections are detectable by definition, but not all detectable sources end up being detected. Indeed, this detectability-conditioned evidence accounts for the fact that extending the model into unobservable regions of the parameters space should not alter its consistency with an observed event.

Putting the above pieces together yields
\begin{align}
\frac
{ \mathcal{Z}(\vec{d}|A,\mathrm{det}) }
{ \mathcal{Z}(\vec{d}|U) }
=
\frac
{ \mathcal{B}_{A/U} }
{ P(\mathrm{det}|A) }
\, ,
\label{eq: evidence ratio det}
\end{align}
which again matches the population-level likelihood but now including selection effects
\citep{
2019PASA...36...10T,
2019MNRAS.486.1086M,
2022hgwa.bookE..45V}.
Note here that as we have used
$P(\mathrm{det}|\vec{d})=1$, $\vec{d}$
represents a particular realization of the data rather than a variable.
As such, the evidences in Eq.~(\ref{eq: evidence ratio det}) are evaluations of the marginal data likelihoods and are not normalized distributions.

We again refer the reader to Appendix~\ref{sec: Practical computation} which describes how to compute the quantities in Eq.~(\ref{eq: evidence ratio det}) in practice.

Similar to Sec.~\ref{sec: Posterior odds}, we can compare two astrophysically-informed models $A$ and $B$, neither of which we have posteriors for, with
\begin{align}
\mathcal{D}_{A/B}
=
\frac
{ \mathcal{Z}(\vec{d}|A,\mathrm{det}) }
{ \mathcal{Z}(\vec{d}|B,\mathrm{det}) }
=
\frac
{ P(\mathrm{det}|B) }
{ P(\mathrm{det}|A) }
\mathcal{B}_{A/B}
\, .
\label{eq: Bayes factor det ratio}
\end{align}
Under equal prior odds, this represents our a posteriori preference for model $A$ over $B$ in producing the data with a detectable GW signal.
Under unequal prior odds, $\mathcal{D}_{A/B}$ should be multiplied by the prior ratio, as in Eq.~(\ref{eq: posterior odds}).

\section{Examples in gravitational-wave astronomy}
\label{sec: Examples in gravitational-wave astronomy}

We now apply our model-comparison formalism to a few example populations and GW events.
We use public GW data from O3
\citep{2023ApJS..267...29A},
in particular the reanalysis of
\cite{2021arXiv210801045T}.
We use posterior samples generated with the \textsc{IMRPhenomXPHM} waveform model
\citep{2021PhRvD.103j4056P} that have been reweighted to a redshift prior corresponding to a uniform source-frame volumetric merger rate.

\subsection{Three-body interactions in young star clusters}
\label{sec: Three-body interactions in young star clusters}

\cite{2021MNRAS.508.3045D} explore
the possibility of forming stellar-mass binary BH mergers through three-body dynamics and assess whether this formation channel can produce mergers whose source properties are consistent with those of GW190521.
They performed $2\times10^5$ direct $N$-body simulations of binary--single interactions including post-Newtonian terms up to the 2.5 order
\citep{
2004PhRvD..70j4011M,
2008AJ....135.2398M}
and a prescription for relativistic kicks
\citep{2018PhRvD..97h4002H}.
The BH population is generated from the simulations of young star clusters (YSCs) by
\cite{2019MNRAS.487.2947D},
including BHs in the pair-instability mass gap between 60--120~$M_\odot$
\citep{2017ApJ...836..244W}
formed via stellar collisions.

The mergers resulting from these dynamical interactions can occur through two main processes: flyby events or exchanges.
In flybys, a tertiary perturber trades energy and momentum with the original binary, leaving the system configuration unchanged.
In exchanges, one component of the original binary is expelled and replaced by the perturber during the interaction.
We label the three possible scenarios as primary--secondary ($m_1$--$m_2$), primary--tertiary ($m_1$--$m_3$), and secondary--tertiary ($m_2$--$m_3$) mergers; i.e., the label indicates the BH components of the final merger.

A small fraction ($\sim0.1\%$) of the YSC mergers contain a BH remnant from a previous merger.
We exclude these sources since their low number would give unreliable statistical estimates.
This results in a total population of 7177 binaries.
We neglect the inclusion of BH spins of the YSC binaries.
\cite{2021MNRAS.508.3045D}
investigated the effect of spins
in post-processing by assuming various Maxwellian distributions for the spin magnitudes, while keeping their directions isotropic.
We reflect the uncertainty -- due to our (lack of) understanding of, e.g., core--envelope coupling, tides, and accretion
\citep{
2018A&A...616A..28Q,
2019ApJ...881L...1F,
2021PhRvD.103f3032S,
2021ApJ...921L...2O,
2022ApJ...930...26S,
2022ApJ...933...86Z,
2023ApJ...952...53M}
-- in the astrophysical origin of spin
\citep{
2022ApJ...937L..13C,
2022MNRAS.517.2738M,
2022PhRvD.106j3019T,
2023arXiv230101312P}
or lack thereof
\citep{
2021PhRvD.104h3010R,
2021ApJ...921L..15G}
by taking the astrophysical spin prior to match the original PE prior (uniform up to maximum magnitudes and isotropic in direction); see Appendix~\ref{sec: Missing and transformed parameters}.

\begin{figure}
\includegraphics[width=\columnwidth]{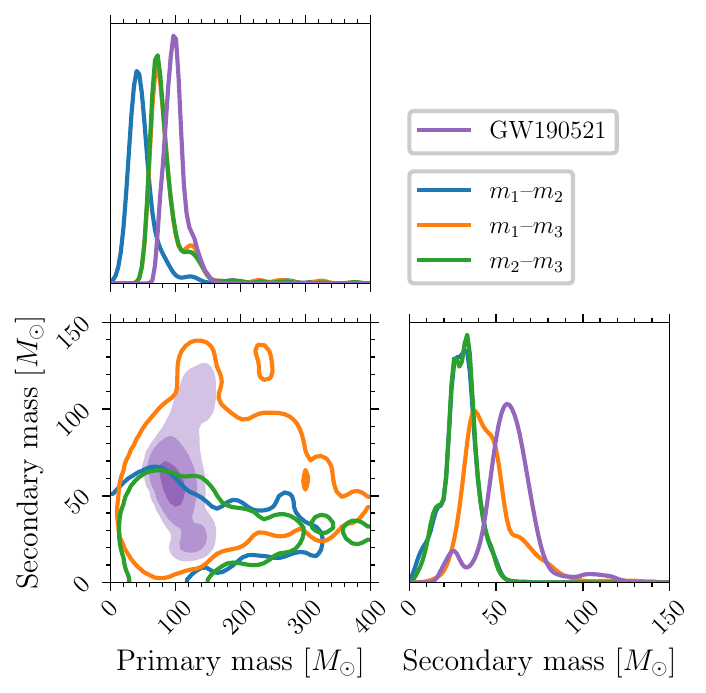}
\caption{
Distributions of the source-frame parimary and secondary component masses of YSC mergers, classified by the type of dynamical interaction: tertiary flyby ($m_1$--$m_2$, blue), original secondary exchanged ($m_1$--$m_3$, orange), and original primary exchanged ($m_2$--$m_3$, green).
Overplotted is the posterior for GW190521 (purple).
The panels on the diagonal present the 1D marginal distributions.
The lower-left panel presents the joint distribution of both masses; for the YSC simulations the contours contain 99\% of sources, while for GW190521 the shading represents darker to lighter the 50\%, 90\%, and 99\% credible regions.
}
\label{fig: ysc}
\end{figure}

In Fig.~\ref{fig: ysc}, we show the distributions of source-frame component masses of mergers from the YSC simulations, classified according to the type of dynamical interaction.
Overplotted is the GW190521 posterior.
Simply by visual inspection it is clear that mergers in which the lighter component of the original binary is exchanged in the dynamical encounter ($m_1$--$m_3$) reach higher masses.
This YSC subpopulation is therefore most likely to accommodate the high masses of an event like GW190521 -- $m_1=98_{-22}^{+34}M_\odot$ and $m_2=57_{-30}^{+27}M_\odot$ (medians and symmetric 90\% CIs) -- compared to $m_1$--$m_2$ and $m_2$--$m_3$ binaries, as pointed out in by
\cite{2021MNRAS.508.3045D}.
Flyby encounters ($m_1$--$m_2$) produce binaries with the lowest total masses and thus the lowest overlap with the mass posterior of GW190521 because the tertiary masses in this model are likely higher than the secondary masses
\citep{2021MNRAS.508.3045D}.

We formalize these conclusions with the Bayes factors listed in Table~\ref{tab: ysc}.
Remembering that these are relative measures, we quantify the likelihood that each dynamical encounter is preferred as a model for GW190521 over the $m_1$--$m_2$ flyby channel using
Eqs.~(\ref{eq: Bayes factor ratio}, \ref{eq: Bayes factor det ratio}); see also Eqs.~(\ref{eq: log Bayes factor ratio mc}, \ref{eq: log Bayes factor det ratio mc}).
Indeed, in support of the qualitative discussion above, we find that dynamical exchanges in which the original primary forms a merging binary with the interacting tertiary are favoured over tertiary flybys with $\ln\mathcal{B}=2.8$ -- a moderately strong but not decisive conclusion.
Secondary--tertiary mergers are also preferred over primary--secondary mergers but only with mild evidence
($\ln\mathcal{B}=1.2$).
These results are in broad agreement with
\cite{2021MNRAS.508.3045D}
(see their Table~2), but are now consistent with the Bayesian PE framework.

\begin{table}
\centering
\renewcommand{\arraystretch}{1.3}
\setlength{\tabcolsep}{10pt}
\begin{tabular}{c|cc}
\hline
Merger & $\ln\mathcal{B}$ & $\ln\mathcal{D}$
\\ \hline
$m_1$--$m_2$ & 0 & 0
\\
$m_1$--$m_3$ & 2.8 & 2.1
\\ 
$m_2$--$m_3$ & 1.2 & 1.1
\\
\hline
\end{tabular}
\caption{
Astrophysical ($\mathcal{B}$) and detection-conditioned ($\mathcal{D}$) Bayes factors comparing subpopulations of mergers within YSCs for GW190521.
Dynamical flybys are labelled $m_1$--$m_2$ and exchanges in which the original binary BH secondary (primary) is replaced with the tertiary BH is labelled $m_1$--$m_3$ ($m_2$--$m_3$).
Values are quoted relative to the $m_1$--$m_2$ sub-channel.
}
\label{tab: ysc}
\end{table}

We compute the detection-conditioned Bayes factors by averaging over $10^5$ draws from the priors (equal to the original PE priors) assumed over the unmodelled parameters (spins, redshifts, and other extrinsic parameters) for each simulated merger, as in Appendix~\ref{sec: Detectable fractions}.
The inclusion of selection effects via Eq.~(\ref{eq: Bayes factor det ratio}) reduces the strength of model preference but does not change the qualitative results (see Table~\ref{tab: ysc}); the distribution of detectable $m_1$--$m_3$ ($m_1$--$m_2$) mergers is most (least) favoured, but only with mild evidence since $|\ln\mathcal{D}|\lesssim2$ in all cases.

\subsection{Hierarchical mergers}
\label{sec: Hierarchical mergers}

Hierarchical BH mergers
\citep{2021NatAs...5..749G}
-- in which at least one binary component is the remnant of a previous merger (or multiple previous mergers) -- may contribute to the binary BH merger rate observed by LIGO/Virgo
\citep{
2021ApJ...915L..35K,
2022PhRvD.106j3013M,
2022ApJ...941L..39W,
2023arXiv230302973L}.
This scenario crucially depends on the properties of the environments in which so-called higher-generation binaries may form
\citep{2021MNRAS.505..339M}
and in particular on their escape speeds, as post-merger remnants receive GW recoils
\citep{1983MNRAS.203.1049F}
that can eject them their hosts and prevent repeated mergers.
If retained, hierarchical mergers result in BHs with distinctively large masses and spins
\citep{2021NatAs...5..749G}.
To a lesser degree, asymmetric masses form through inter-generational pairings, though even cluster dynamics tend to preferentially lead to nearly equal-mass systems
\citep{2016ApJ...824L..12O, 2019MNRAS.487.2947D, 2020MNRAS.498..495D, 2023MNRAS.522..466A}
due to, e.g., mass segregation.
The hierarchical merger scenario has therefore been invoked as a possible origin for the formations of GW190412
\citep{
2020PhRvD.102d3015A,
2020PhRvL.125j1103G,
2020ApJ...896L..10R,
2021ApJ...908..194T,
2021MNRAS.502.2049L}
and GW190521
\citep{
2020ApJ...900L..13A,
2020ApJ...902L..26F,
2021MNRAS.508.3045D,
2021ApJ...915L..35K,
2021ApJ...920..128A,
2021ApJ...908..194T,
2021MNRAS.502.2049L}.

Labelling BHs that formed as the endpoint of stellar evolution as ``first-generation'' (1G) BHs and the merger product of two 1G BHs as a ``second-generation'' (2G) BH, we test which merger generations are most consistent with the GW events GW190412 and GW190521 using a simple model of hierarchical BH formation
\citep{
2023arXiv230504987G,
matthew_mould_2023_7807210}.
In this model we define a ``cluster'' as a collection of BHs that can each pair to form a merging binary hosted in an environment with a common escape speed $v_\mathrm{esc}$.
We generate $10^3$ distinct clusters whose escape speeds are chosen uniform randomly between $0\,\mathrm{km/s}$ and $10^3\,\mathrm{km/s}$, each with an initial distribution of $10^4$ 1G BHs whose masses are drawn from a powerlaw between $5M_\odot$ and $50M_\odot$ (reflecting cutoffs due to pair-instability supernovae) with index $-2$ (roughly tracing the stellar initial mass function) and whose spins have uniform magnitudes between 0 and 0.2 (assuming low natal spins).
One by one, BHs are drawn and paired with a less massive binary partner, both with uniform probabilities.
Their spin directions are sampled isotropically.
The properties of the merger remnant -- its mass, spin magnitude, and kick magnitude -- are then estimated
\citep{2016PhRvD..93l4066G}.
If the kick exceeds the assigned escape speed of the host cluster the remnant is ejected, but otherwise it is retained.
The steps above are then iterated to exhaustion of the available BHs and this is repeated for all the clusters.
Finally, the mergers from all $10^3$ clusters are concatenated to form one population containing $\sim7\times10^6$ mergers.

This simple procedure is able to rapidly generate large populations of hierarchical mergers.
However, we note that the various assumptions in our parametrization are not necessarily realistic in an astrophysical setting.
However, it is useful in its computational efficiency and produces population distributions reflecting the key features of hierarchical BH formation, serving as an illustrative example for this work.

\begin{figure}
\includegraphics[width=\columnwidth]{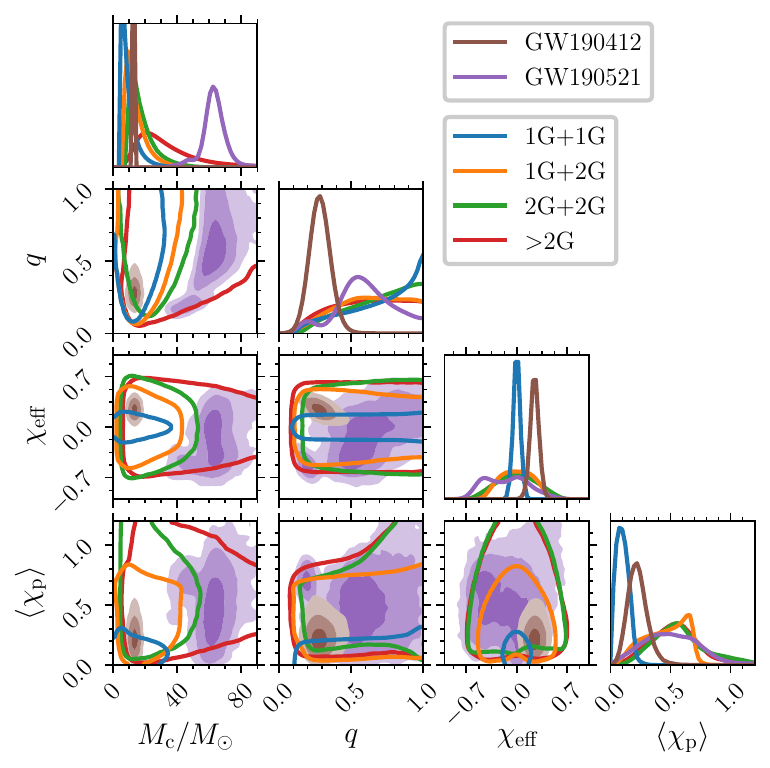}
\caption{
Distributions of chirp mass $M_\mathrm{c}$, mass ratio $q$, effective aligned spin $\chi_\mathrm{eff}$, and averaged effective precessing spin $\langle\chi_\mathrm{p}\rangle$.
Diagonal and lower panels present one- and 2D marginals, respectively.
The 2D shaded regions for GW190412 (brown) and GW190521 (purple) represent the 50\%, 90\%, and 99\% credible regions, from darker to lighter.
The simulated hierarchical mergers are split into subpopultions of binaries containing two 1G BHs (1G$+$1G, blue), a 1G BH and a 1G+1G remnant (1G$+$2G, orange), two 2G BHs (2G$+$2G, green), or at least one BH having undergone more than one previous merger ($>$2G, red).
Their 2D distributions are given by contours containing 99\% of sources.
} 
\label{fig: hierarchical}
\end{figure}

Figure~\ref{fig: hierarchical} displays the output of these simple simulations.
We split the full population into four distinct subsets: those in which the merging binary contained two 1G BHs (1G$+$1G), one 1G BH and one 2G BH (1G$+$2G), two 2G remnants (2G$+$2G), and at least one component which has already undergone more than one previous merger ($>$2G).
We plot the population distributions of the binary chirp mass $M_\mathrm{c}$, mass ratio $q$, effective aligned spin $\chi_\mathrm{eff}$
\citep{2008PhRvD..78d4021R},
and averaged effective precessing spin $\langle\chi_\mathrm{p}\rangle$
\citep{2021PhRvD.103f4067G};
the first three are conserved quantities while $\langle\chi_\mathrm{p}\rangle$ varies over the inspiral (but not over a spin-precession cycle) and is bounded in $[0,2]$.
The effective spins in particular have been shown to present features identifying hierarchical mergers
\citep{
2021MNRAS.505..339M,
2021PhRvD.104h4002B}.
Over plotted are the posteriors of these parameters for GW190412 and GW190521.
The higher-generation distributions unsurprisingly reach higher masses and spins which can accommodate the heavy BHs and possibility of tilted spins in GW190521 and non-zero $\chi_\mathrm{eff}$ of GW190412.
The 1G$+$2G and $>$2G subpopulations overlap most with the marginal mass-ratio distributions of GW190412 as 1G$+$1G and 2G$+$2G formations are more likely to pair equal-mass BHs.

We again formalize these notions with the Bayes factors from Eqs.~(\ref{eq: Bayes factor ratio}, \ref{eq: Bayes factor det ratio}).
In this case, since we have a large number ($\sim10^7$) of total mergers, when computing the detection fractions we assign a single value drawn from the PE prior for the set of unmodelled parameters (redshift, inclination, right ascension, declination, and polarization) to each merger, rather than averaging over a set of Monte Carlo realizations.
We compute astrophysical and detection-conditioned Bayes factors with respect to the 1G$+$1G subpopulation and report the results in Table~\ref{tab: hierarchical}.

\begin{table}
\centering
\renewcommand{\arraystretch}{1.3}
\setlength{\tabcolsep}{10pt}
\begin{tabular}{c|cc|cc}
\hline
& \multicolumn{2}{c|}{GW190412} & \multicolumn{2}{c}{GW190521}
\\
Generation & $\ln\mathcal{B}$ & $\ln\mathcal{D}$ & $\ln\mathcal{B}$ & $\ln\mathcal{D}$
\\ \hline
1G$+$1G & 0 & 0 & 0 & 0
\\
1G$+$2G & 3.2 & 2.3 & 94 & 93
\\
2G$+$2G & 1.2 & $-$0.1 & 98 & 97
\\ 
$>$2G & 1.7 & $-$0.7 & 102 & 100
\\
\hline
\end{tabular}
\caption{
Intrinsic ($\mathcal{B}$) and detectability-conditioned ($\mathcal{D}$) Bayes factors for higher-generation over 1G origins for GW190412 and GW190521.
}
\label{tab: hierarchical}
\end{table}

For GW190412, 1G$+$2G formation is favoured with respect to other hierarchical mergers, with moderate evidence ($\ln\mathcal{B}=3.2$) over first-generation binaries and milder evidence with respect to higher generations.
Owing to the confidently unequal masses ($q=0.28_{-0.07}^{+0.09}$) of GW190412, any population that pairs unequal-mass BHs while allowing for its positively aligned spin ($\chi_\mathrm{eff}=0.25_{-0.10}^{+0.10}$) and precise chirp mass ($M_\mathrm{c}=13.3_{-0.4}^{+0.5}M_\odot$) will be preferred.
In particular, the mass-ratio distributions of 1G$+$2G and $>$2G binaries are flatter compared to equal-generation pairings, while 1G$+$2G binaries are more densely located in the region of GW190412's chirp mass (see Fig.~\ref{fig: hierarchical}).
The detectable populations of 2G$+$2G and $>$2G binaries become disfavoured with respect to 1G$+$1G binaries in describing GW190412 with $|\ln\mathcal{D}|\lesssim1$, while the 1G$+$2G subpopulation remains favoured, though more weakly.

In the case of GW190521, we note that 1G$+$1G binaries often do not overlap with the GW posterior due to the maximum 1G mass $m_\mathrm{max}=50M_\odot$.
This means that, in computing the Bayes factors in Eqs.~(\ref{eq: log Bayes factor ratio mc}, \ref{eq: log Bayes factor det ratio mc}), the result becomes sensitive to the number of posterior samples used due to low effective counts; increasing the posterior sample size likely increases the denominator of Eq.~(\ref{eq: Bayes factor mc}) without increasing the numerator.
We use all the samples available in the public data ($\sim2\times10^5$ for GW190412 and $\sim2\times10^4$ for GW190521),
which implies that the Bayes factors for GW190521 in Table~\ref{tab: hierarchical} represent lower limits.
The result, however, is clear: in our simplified model of hierarchical mergers with our selected parametrization, GW190521 is confidently favoured to contain a higher-generation BH over our assumed 1G population, with $\ln\mathcal{B}\approx100$.
In fact, the subpopulation of binaries in which at least one BH has undergone more than one merger ($>$2G) is most favoured, confidently ($\ln\mathcal{B}>5$) over 1G$+$2G binaries but less so ($\ln\mathcal{B}<5$) over 2G$+$2G binaries.
This conclusion, or its strength, would change if, e.g., we increased the maximum allowed mass of 1G BHs in our parametrization.
Including selection effects, we find these results are almost unchanged for the detectable hierarchical-merger distributions (see Table~\ref{tab: hierarchical}).

\begin{figure}
\includegraphics[width=\columnwidth]{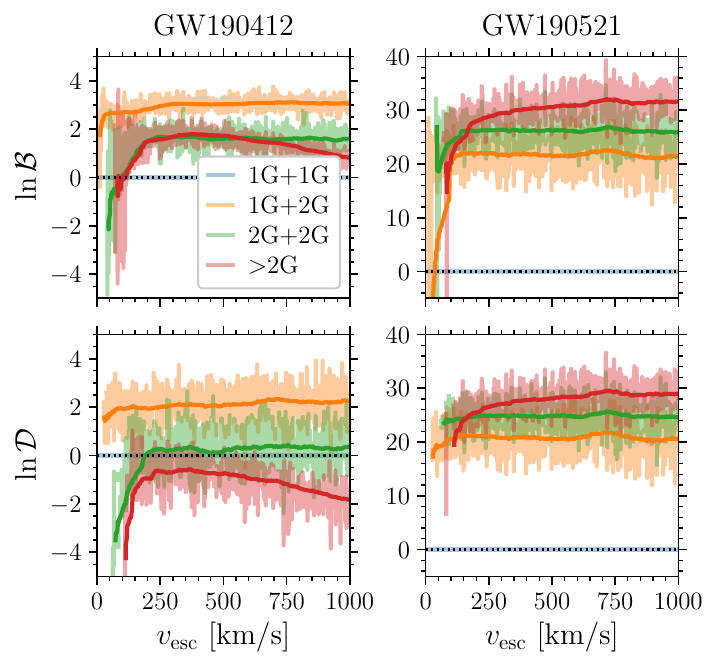}
\caption{
Intrinsic ($\mathcal{B}$, top row) and detectability-conditioned ($\mathcal{D}$, bottom row) Bayes factors for higher-generation origins of GW190412 (left column) and GW190521 (right column) as a function of the host cluster escape speed $v_\mathrm{vesc}$.
The Bayes factors for binaries with one 1G and one 2G BH (1G$+$2G, orange), two 2G BHs (2G$+$2G, green), and at least one BH of even higher generation ($>$2G, red) are given with respect to 1G$+$1G binaries (blue).
Shaded lines represent the Monte Carlo estimation while solid lines give a moving average.
The horizontal dotted black lines mark the boundaries of equal preference.
}
\label{fig: vesc}
\end{figure}

Furthermore, we compute the Bayes factors for the populations of both intrinsic and detectable mergers \emph{within} each cluster -- i.e., as a function of the host escape speed $v_\mathrm{vesc}$ -- over the binary merger generations.
The results are shown in Fig.~\ref{fig: vesc}.
We first note that the noise is due to the fact that each cluster produces a number of mergers that is at most one less than the initial number ($10^4$) of 1G BHs and reflects the statistical uncertainty in the Monte Carlo approximation.
At low escape speeds there are not many -- if any -- higher-generation mergers, implying the approximated Bayes factors (see Appendix~\ref{sec: Practical computation}) cannot be trusted there.

For GW190412, $>$2G binaries become disfavoured relative to other higher-generation populations at larger escape speeds because such clusters more efficiently retain larger masses that are inconsistent with those of the observed event; for this same reason, in the case of GW190521 the astrophysical and detectable $>$2G Bayes factors increase on average with larger cluster escape speeds.
The astrophysical (detectable) $>$2G subpopulation remains favoured (becomes disfavoured) over the 1G$+$1G subpopulation for increasing $v_\mathrm{vesc}$, while the Bayes factors for 1G$+$2G and 2G$+$2G across all clusters are consistent with the overall population in Table~\ref{tab: hierarchical}.
On the other hand, for GW190521 the per-cluster preference for higher-generation binaries is reduced compared to the overall population, but there is still a decisive conclusion against first-generation binaries.
We reiterate that this conclusion depends on our choice of parametrization.

\section{Conclusions}
\label{sec: Conclusions}

Even as the number of detected GW events grows into the triple digits, there are individual detections which are interesting on their own merit.
They may have unusual source properties that make them outliers with respect to the rest of the inferred population.
For these detections we may wish to individually assess the degree to which they are reproducible by a given formation scenario.
Some previous analyses have employed the heuristic fraction of synthetic mergers from a simulated population that lie within a chosen volume of the measured parameters of a real GW event.
This metric is unsuitable as it does not account for the full Bayesian posterior and prior information in the original PE, it depends on a heuristically constructed confidence region, and it fails in the limit of precise measurements (Sec.~\ref{sec: Heuristic approximation}).

We point out that a statistically consistent replacement is given by the posterior odds (Secs.~\ref{sec: Bayes factors} and \ref{sec: Posterior odds}), with a tractable computation in practice (Appendix~\ref{sec: Practical computation}).
We also show how the influence of selection biases can be accommodated in this statistical framework (Sec.~\ref{sec: Astrophysical populations and selection effects}), consistent with the usual hierarchical GW population likelihood.
In fact, rather than starting from a single GW event and working up, one can similarly derive Eqs.~(\ref{eq: evidence ratio det}, \ref{eq: Bayes factor det ratio}) starting from the full population likelihood and working down by considering the limiting case of a single observation and fixed population models.

Crucially, the posterior odds is a relative or comparative measure; one can only test whether a particular astrophysical model is more likely to produce mergers consistent with a given GW event compared to another, but not whether the model is good in absolute terms -- a natural outcome of working within the Bayesian framework.
We used this fact to assess the degree to which selected GW observations are better explained by various simulated formation pathways relative to one another (Sec.~\ref{sec: Examples in gravitational-wave astronomy}).
In particular, we demonstrated comparisons between subpopulations (e.g., first-generation against second-generation mergers) within a particular formation environment (e.g., dynamical clusters).

As mentioned in Sec.~\ref{sec: Practical summary}, the calculation we presented is actually a two-step procedure, the second of which we derived in this work: first, one selects an event of interest, then reanalyses it in the context of competing population models to deduce which better describes the data.
We neglected a quantitative inclusion of the first step as we were interested in the information provided by a single event irrespective of other events in the combined observational catalogue.
One can account for the information provided by the remaining catalogue, assuming that the single-event selection is equally probable among all events, with population-informed priors
\citep{2021PhRvD.104h3008M}.
Otherwise, the selection of that event must be accounted for to avoid biases, akin to those due to search sensitivity which results in the additional detection-probability factors in Eq.~(\ref{eq: evidence det}).
To take this approach, one can follow \cite{2022ApJ...926...34E} who derive the population-level likelihood for a catalogue of detections but marginalise over a single event that may be an outlier while self-consistently accounting for the very selection of that event.
The problem tackled here is the converse: one analyses a single interesting event while marginalising over the remaining catalogue, accounting for the same selection effect just mentioned; cf. their Eqs. (7) and (10).
Defining an event as interesting may depend on the remaining catalogue -- e.g., it is clearly separated in parameter space from all others -- or on astrophysical prior knowledge -- e.g., it lies in a predefined data-independent mass gap.

This framework can be used not just when selecting a single event from a larger catalogue of observations, but also when there is only a single observation available in the first place. This may be the case for LISA
\citep{2017arXiv170200786A},
where the most pessimistic forecasts predict just $\sim\mathcal{O}(1)$ massive BH merger and extreme mass-ratio inspiral observations annually
\citep{2023LRR....26....2A}.
The degree to which we can conclude model preference in the presence of observational biases in this scenario of limited available data will therefore be an interesting topic for future work.

\section*{Acknowledgements}

We thank Christopher Moore, Riccardo Buscicchio, Floor Broekgaarden, and Christopher Berry for helpful discussions.
M.Mo. and D.G. are supported by European Union's H2020 ERC Starting Grant No. 945155--GWmining, Cariplo Foundation Grant No. 2021-0555, and the ICSC National Research Centre funded by NextGenerationEU. D.G. is supported by Leverhulme Trust Grant No. RPG-2019-350. 
M.D. is supported by Cariparo Foundation Grant. No. 55440.
M.Ma. is supported by European Union’s H2020 ERC Consolidator Grant No. 770017–DEMOBLACK and by the German Excellence Strategy via the Heidelberg Cluster of Excellence (EXC 2181 - 390900948) STRUCTURES.

\section*{Data Availability}

The data underlying this article will be shared on reasonable request to the corresponding author.
We make available a reproducible tool to perform the computation along with examples at
\href
{https://github.com/mdmould/popodds}
{github.com/mdmould/popodds};
\citealt{matthew_mould_2023_7647704}.

\bibliographystyle{mnras_tex_edited}
\bibliography{draft}

\appendix

\section{Practical computation}
\label{sec: Practical computation}

\subsection{Bayes factor}

The integral in the Bayes factor of Eq.~(\ref{eq: Bayes factor}) can be computed as a Monte Carlo summation in two ways that are equivalent (modulo numerical issues -- see below), unlike for the heuristic fraction:
\begin{align}
\mathcal{B}_{A/U}
\approx
\frac{1}{N}
\sum_{i=1}^N
\frac
{ \pi(\vec{\theta}_i|A) }
{ \pi(\vec{\theta}_i|U) }
\approx
\frac{1}{N_A}
\sum_{j=1}^{N_A}
\frac
{ p(\vec{\theta}_{Aj}|\vec{d},U) }
{ \pi(\vec{\theta}_{Aj}|U) }
\, .
\label{eq: Bayes factor mc}
\end{align}
The first expression can be computed if one has posterior samples $\{\vec{\theta}_i\}_{i=1}^N \sim p(\vec{\theta}|\vec{d},U)$ for a given GW event and can evaluate the population density $\pi(\vec{\theta}|A)$.
The second expression can be computed if one can evaluate the posterior density $p(\vec{\theta}|\vec{d},U)$ of the GW event and has synthetic sources $\{\vec{\theta}_{Aj}\}_{j=1}^{N_A} \sim \pi(\vec{\theta}|A)$ simulated from an astrophysical population.
In either case we must be able to evaluate the prior density $\pi(\vec{\theta}|U)$ from the original PE.

Besides this, the better expression to compute is the one in which the additional uncertainty introduced by the integral approximation is lower.
This depends on the variances of the GW posterior and simulated population, and on the number of samples over which the summation is taken
\citep{2023arXiv230406138T}.
If one already has an explicit form for $\pi(\vec{\theta}|A)$ given by, e.g., a parametrized model, the first expression of Eq.~(\ref{eq: Bayes factor mc}) should be used.
Otherwise, one should use the first expression of Eq.~(\ref{eq: Bayes factor mc}) if the error on the GW event posterior $p(\vec{\theta}|\vec{d},U)$ is smaller than the range of the astrophysical population $\pi(\vec{\theta}|A)$, and vice versa.
The former is most often true in the case of GW data and as such is our default choice (e.g., a single binary mass measurement is more narrow than the entire range of astrophysical masses, though this is not necessarily true for parameters that are typically not well measured such as BH spins).

It may be the case that we do not have access to a closed-form expression for the uninformative PE prior $\pi(\vec{\theta}|U)$ but do have samples drawn from it -- in this case a density estimator is required.
If building an informative prior from astrophysical simulations, we also need to construct an estimator for $\pi(\vec{\theta}|A)$ from the finite set of simulated sources.
For the examples following in Sec.~\ref{sec: Examples in gravitational-wave astronomy} we use Gaussian kernel density estimates (KDEs)
\citep{
2015mdet.book.....S,
2020SciPy-NMeth};
for more complex or higher-dimensional distributions other density estimation techniques may be more accurate, e.g., normalizing flows
\citep{2019arXiv190809257K, 2019arXiv191202762P}.
On the other hand, if a continuous representation of the posterior density $p(\boldsymbol{\theta}|\boldsymbol{d},U)$ is already available (e.g., through neural posterior estimation as in
\citealt{
Green_2021,
2021PhRvL.127x1103D}),
the discrete simulated sources can be used directly.

It should be noted, however, that the expression on the right of Eq.~(\ref{eq: Bayes factor mc}) has a numerical divergence if $\pi(\vec{\theta}|A)$ has support beyond the support of $\pi(\vec{\theta}|U)$ -- i.e., there may be simulated sources $\vec{\theta}_{Aj}$ for which $\pi(\vec{\theta}_{Aj}|U)=0$, whereas the equivalent integrand $\mathcal{L}(\vec{d}|\vec{\theta}_{Aj}) / \mathcal{Z}(\vec{d}|U)$ is perfectly well defined.
This does not occur in the left-hand expression, for which we are guaranteed that $\pi(\vec{\theta}_i|U)>0$ by definition.

\subsection{Detectable fractions}
\label{sec: Detectable fractions}

We also need to compute the detectable fractions in Eq.~(\ref{eq: Bayes factor det ratio}).
In principle these quantities must be estimated by injecting signals with source properties $\vec{\theta}$ into the full detection pipeline, with $p(\mathrm{det}|\vec{\theta}) = 1$ $(0)$ for signals which are (not) recovered.
However, this presents prohibitive computational expense when required for large populations.
Additionally, the detection probability depends on all the source properties -- both intrinsic and extrinsic -- but we might only be interested in modelling the population of the former, or even just a subset of them.
Each source must be assigned single values for the unmodelled parameters -- which increases the uncertainty in the population-averaged detectability -- or these parameters must be manually marginalized over -- a procedure which must be done numerically and increases the computational burden.

In practice, a source is often approximated as detectable if its SNR is above a chosen threshold value.
And recent approaches have utilized machine learning to enable efficient evaluations of source detectability
\citep{
2020PhRvD.102j3020G, 2020arXiv200710350W,
2022ApJ...927...76T, 2023MNRAS.522.6043C}.
The process of generating a waveform, projecting the signal into the detector frame, computing the detection statistic (e.g., SNR), and evaluating the detectability can be reduced to the evaluation of a neural network trained on sources already assessed in the full detection pipeline.
In particular, such approximants can process large batches of inputs very efficiently.
They are therefore well suited for averaging detection probabilities over large populations of simulated sources and, furthermore, for handling unmodelled parameters that require an additional averaging step, as mentioned above.
We therefore choose to use the neural network implementation of
\cite{2020PhRvD.102j3020G},
assuming a three-detector LIGO/Virgo network with sensitivities corresponding to the third observing run (O3) and a network SNR threshold of 12
\citep{2020LRR....23....3A}.
For each binary, the network takes as input ($\vec{\theta}$) the detector-frame component masses, spin vectors, redshift, orbital inclination, right ascension, declination, and GW polarization, and outputs the classification $p(\mathrm{det}|\vec{\theta})=1$ (0) for sources which are (not) detectable.

When the detection probabilities of each simulated binary are estimated directly, whether in this way or with a different pipeline, the detection fractions required in Eq.~(\ref{eq: Bayes factor det ratio}) can then be computed as expectation values over the population.
For our set of simulated mergers
$\{\vec{\theta}_{Aj}\}_{j=1}^{N_A} \sim \pi(\vec{\theta}|A)$,
Eq.~(\ref{eq: Pdet population}) can be approximated with
\begin{align}
P(\mathrm{det}|A)
\approx
\frac{1}{N_A} \sum_{j=1}^{N_A} P(\mathrm{det}|\vec{\theta}_{Aj})
\, .
\end{align}

\subsection{Numerical implementation}

Putting everything together to compare the GW event posterior $\{\vec{\theta}_i\}_{i=1}^N \sim p(\vec{\theta}|\vec{d},U)$ to two astrophysical simulations $\{\vec{\theta}_{Aj}\}_{j=1}^{N_A} \sim \pi(\vec{\theta}|A)$ and $\{\vec{\theta}_{Bk}\}_{k=1}^{N_B} \sim \pi(\vec{\theta}|B)$, we get for the Bayes factor between the intrinsic populations
\begin{align}
\mathcal{B}_{A/B}
\approx
\frac
{ \sum_{i=1}^N \pi(\vec{\theta}_i|A) / \pi(\vec{\theta}_i|U) }
{ \sum_{n=1}^N \pi(\vec{\theta}_i|B) / \pi(\vec{\theta}_i|U) }
\, ,
\label{eq: Bayes factor ratio mc}
\end{align}
and between the detectable populations
\begin{align}
\mathcal{D}_{A/B}
\approx
\frac
{ N_A \sum_{k=1}^{N_B} P(\mathrm{det}|\vec{\theta}_{Bk}) }
{ N_B \sum_{j=1}^{N_A} P(\mathrm{det}|\vec{\theta}_{Aj}) }
\mathcal{B}_{A/B}
\, .
\label{eq: Bayes factor det ratio mc}
\end{align}

It may be more numerically stable to evaluate the logarithms of the probability densities in Eq.~(\ref{eq: Bayes factor ratio mc}) and then compute the logarithms of the integrals using a log-sum-exp implementation (our default).
The Bayes factors between the two populations are then given by
\begin{align}
\ln \mathcal{B}_{A/B}
\approx\,&
\textstyle
\ln \sum_{i=1}^N \exp
[ \ln \pi(\vec{\theta}_i|A) - \ln \pi(\vec{\theta}_i|U) ]
\nonumber \\
-\,&
\textstyle
\ln \sum_{i=1}^N \exp
[ \ln \pi(\vec{\theta}_i|B) - \ln \pi(\vec{\theta}_i|U) ]
\, ,
\label{eq: log Bayes factor ratio mc}
\\
\ln \mathcal{D}_{A/B}
=\,&
\ln \mathcal{B}_{A/B} - \ln P(\mathrm{det}|A) + \ln P(\mathrm{det}|B)
\, .
\label{eq: log Bayes factor det ratio mc}
\end{align}

\section{Missing and transformed parameters}
\label{sec: Missing and transformed parameters}

In many situations, the new prior may model a subset or transformed versions of the original parameters, or both.
An example from GW astronomy is that, while $\vec{\theta}$ contains the detector-frame binary masses, component spins, luminosity distance, and sky location, etc., a model of compact-binary formation (such as a population-synthesis simulation) may only model source-frame masses and redshift, without modelling the other extrinsic parameters or perhaps the spins.
These differences enter the calculation of the Bayes factor.

\subsection{Missing parameters}
\label{sec: Missing parameters}

Suppose that the astrophysical prior only models a subset $\vec{\phi}$ (e.g., intrinsic binary parameters) of the parameters $\vec{\theta} = (\vec{\phi},\vec{\psi})$, where $\vec{\psi}$ are the unmodeled parameters (e.g., extrinsic binary parameters).
One cannot simply replace $\vec{\theta}$ with $\vec{\phi}$ in the integral of Eq.~(\ref{eq: Bayes factor}).
Since $\vec{\phi}$ and $\vec{\psi}$ are independent of each other in our astrophysical prior and if we take equal priors $\pi(\vec{\psi}|A) = \pi(\vec{\psi}|U)$ (as is usual in population inference; \citealt{2022hgwa.bookE..45V}), then
\begin{align}
\mathcal{B}_{A/U}
=
\int \pi(\vec{\phi}|A)
\int \frac{ p(\vec{\phi},\vec{\psi}|\vec{d},U) } { \pi(\vec{\phi}|\vec{\psi},U) }
\, \dd\vec{\psi} \, \dd\vec{\phi} 
\, .
\label{eq: missing parameters}
\end{align}
Only if the original PE prior over $\vec{\phi}$ is independent of $\vec{\psi}$, i.e.,
$\pi(\vec{\phi}|\vec{\psi},U) = \pi(\vec{\phi}|U)$,
may we compute the Bayes factor by replacing $\vec{\theta}$ with $\vec{\phi}$; this may be true in some cases but is not in general.
For example, consider a model of source-frame binary masses when the original prior was placed over detector-frame masses;
here, one must account for the redshift dependence.

\subsection{Transformed parameters}
\label{sec: Transformed parameters}

When dealing with a such transformation $\tilde{\vec{\theta}} = \tilde{\vec{\theta}}(\vec{\theta})$ (e.g., source-frame masses and redshift) of the original parameters $\vec{\theta}$ (e.g., detector-frame masses and luminosity distance), one can either convert the original posterior $p(\vec{\theta}|\vec{d},U)$ and prior $\pi(\vec{\theta}|U)$ to distributions over the new prior parameters $\tilde{\vec{\theta}}$, or convert the new prior $\pi(\tilde{\vec{\theta}}|A)$ in terms of the original parameters $\vec{\theta}$.
The corresponding transformation of the distributions is
$p(\tilde{\vec{\theta}}) = p(\vec{\theta}) |d\vec{\theta} / d\tilde{\vec{\theta}}|$,
where the scaling is given by the Jacobian determinant of the inverse transformation.
In fact, either parametrization is equivalent because
\begin{align}
\int p(\vec{\theta}|\vec{d},U) \frac { \pi(\vec{\theta}|A) } { \pi(\vec{\theta}|U) } \dd\vec{\theta}
=
\int p(\tilde{\vec{\theta}}|\vec{d},U) \frac{ \pi(\tilde{\vec{\theta}}|A) } { \pi(\tilde{\vec{\theta}}|U) } \dd\tilde{\vec{\theta}}
\, .
\end{align}

Given the original posterior samples $\vec{\theta} \sim p(\vec{\theta}|\vec{d},U)$, one can readily construct the transformed posterior samples as $\tilde{\vec{\theta}}(\vec{\theta}) \sim p(\tilde{\vec{\theta}}|\vec{d},U)$.
The Bayes factor from Eq.~(\ref{eq: Bayes factor}) may be written as
\begin{align}
\mathcal{B}_{A/U}
&=
\int p(\vec{\theta}|\vec{d},U)
\frac{ \pi\big( \tilde{\vec{\theta}}(\vec{\theta})|A \big) } { \pi(\vec{\theta}|U) }
\left| \frac{\dd\tilde{\vec{\theta}}}{\dd\vec{\theta}} \right|
\dd\vec{\theta}
\\ &=
\int p(\tilde{\vec{\theta}}|\vec{d},U)
\frac{ \pi(\tilde{\vec{\theta}}|A) } { \pi\big( \vec{\theta}(\tilde{\vec{\theta}})|U \big) }
\left| \frac{\dd\tilde{\vec{\theta}}}{\dd\vec{\theta}} \right|
\dd\tilde{\vec{\theta}}
\, ,
\label{eq: transformed parameters}
\end{align}
where we have written the first expression in terms of the original parameters $\vec{\theta}$ but kept the new prior model $A$ in terms of its parameters $\tilde{\vec{\theta}}$, and we have written the second expression in terms of the transformed parameters $\tilde{\vec{\theta}}$ but kept the original prior $U$ in terms of its parameters $\vec{\theta}$.

As a concrete (and rather common -- see, e.g., \citealt{2019PhRvX...9c1040A, 2021PhRvX..11b1053A, 2021arXiv210801045T, 2021arXiv211103606T}) example, consider setting a prior over the redshifted component masses $m_{1z}=(1+z)m_1$ and $m_{2z}=(1+z)m_2$ of a GW event as observed in the detector frame and then modelling the source-frame masses $m_1 \geq m_2$ in the population without modelling the redshift $z$.
As usual, we assume priors over the remaining binary parameters that are independent of masses and redshift, such that the corresponding terms in the Bayes factor cancel, as in Eq.~(\ref{eq: missing parameters}).
The Jacobian for the transformation $(m_{1z}, m_{2z}, z) \mapsto (m_1, m_2, z)$ is $(1+z)^2$.
In this example, we would therefore have that
\begin{align}
\mathcal{B}_{A/U}
=&
\int \dd m_{1z} \dd m_{2z} \dd z \, (1+z)^2 \,
p ( m_{1z}, m_{2z}, z | \vec{d}, U )
\nonumber \\
&\times
\frac { \pi \big( m_{1z}/(1+z), m_{2z}/(1+z) \big| A \big) } { \pi ( m_{1z}, m_{2z} | U ) }
\nonumber \\
=&
\int \dd m_1 \dd m_2 \dd z \, (1+z)^2 \,
p( m_1, m_2, z | \vec{d}, U )
\nonumber \\
&\times
\frac { \pi ( m_1, m_2 | A) } { \pi \big( m_1(1+z), m_2(1+z) \big| U \big) }
\, ,
\label{eq: redshift transformation}
\end{align}
where one should note the assumed cancellation $\pi(z|A) / \pi(z|U) = 1$.
To reiterate, this is only valid because we have accounted for the transformation of variables.

\section{Simple example}
\label{sec: Simple example}

\subsection{Bayes factor}

We now consider an illustrative univariate example in which a uniform prior ($U$) on a parameter $x$ between $a=0$ and $b=100$ results in a Gaussian posterior at $\hat{x}=50$ with standard deviation $\hat{\sigma}=5$.
We take normal distributions with means $\mu$ and standard deviations $\sigma$ as alternative priors ($A$) and compute the Bayes factor over the uniform prior.
In this simplified case, the Bayes factor has the closed form
\begin{align}
\mathcal{B}_{A/U}
=
\frac{b-a}{\sqrt{2\pi (\hat{\sigma}^2 + \sigma^2)}} \exp\left( -\frac{1}{2} \frac{(\hat{x} - \mu)^2}{\hat{\sigma}^2 + \sigma^2} \right)
\, .
\label{eq: Bayes factor simple}
\end{align}

\begin{figure}
\includegraphics[width=\columnwidth]{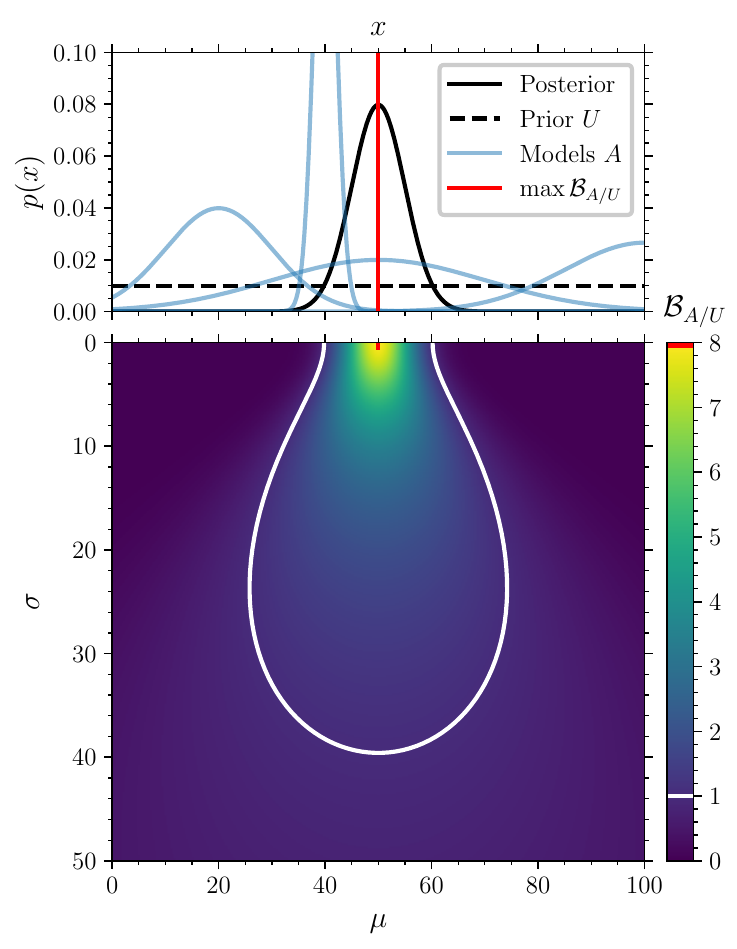}
\caption{
Bayes factors $\mathcal{B}_{A/U}$ in a simple example with a univariate parameter $x$ (Sec.~\ref{sec: Simple example}).
In the top panel we display the fiducial posterior for $x$ (solid black curve), the assumed uniform prior from which this posterior was inferred (dashed black line), and several example Gaussians which act as alternative priors (blue curves).
In red is the delta distribution corresponding to the model with the maximum Bayes factor.
In the bottom panel we display the Bayes factor as a function of the mean and standard deviation of the Gaussian prior, with white marks denoting the equal preference $\mathcal{B}_{A/U}=1$ boundary inside which the alternative prior is favoured.
}
\label{fig: simple}
\end{figure}

In Fig.~\ref{fig: simple} we show the prior, posterior, and a few examples of the fiducial model in the top panel.
The Bayes factor is plotted
as a function of $\mu$ and $\sigma$
in the bottom panel.
Priors which assign more support further from the bulk of the posterior are disfavoured with respect to the original flat prior, meaning
$\mathcal{B}_{A/U}<1$.
In the limit of broad models as $\sigma\to\infty$, the Gaussian priors become disfavoured since their densities are always lower than that of the flat prior.
Models become favoured if well localized and centred on the posterior.
The changeover occurs at $\mathcal{B}_{A/U}=1$; this boundary (see below) is marked in white in Fig.~\ref{fig: simple}, inside (outside) which the Gaussian prior is favoured (disfavoured) over the flat prior.
The maximum Bayes factor occurs for a delta distribution placed at the peak of the posterior
\citep{2023PhRvR...5b3013P},
i.e, for $\mu=\hat{\mu}$ and $\sigma=0$, as indicated in red.

According to Eq.~(\ref{eq: Bayes factor simple}), the $\mathcal{B}_{A/U}=1$ boundary is given by
\begin{align}
\mu
=
\hat{x} \pm \sqrt{ 2(\hat{\sigma}^2+\sigma^2) \ln \frac{ b-a } { \sqrt{ 2\pi (\hat{\sigma}^2 + \sigma^2) } } }
\end{align}
and is marked in white in Fig.~\ref{fig: simple}.
The largest value of $\sigma$ with $\mathcal{B}_{A/U}=1$ occurs when $\sigma = \sqrt{(b-a)^2/(2\pi) - \hat{\sigma}^2} \approx 40$ and $\mu=\hat{x}=50$, while the largest and smallest values of $\mu$ with $\mathcal{B}_{A/U}=1$ occur when $\sigma = \sqrt{(b-a)^2/(2\pi\mathrm{e}) - \hat{\sigma}^2} \approx 24$ and $\mu = \hat{x} \pm (b-a) / \sqrt{2\pi\mathrm{e}} \approx 50 \pm 24$.
The $\mathcal{B}_{A/U}=1$ turning point at $\sigma=0$ gives $\mu\approx40,60$.
The maximum Bayes factor model, indicated in red, is a delta distribution at $x=\hat{x}$
\citep{2023PhRvR...5b3013P}
with $\max\mathcal{B}_{A/U} = (b-a) / (\sqrt{2\pi} \hat{\sigma}) \approx 8$.

Note that, though our assumed posterior has support outside the prior range (which in reality is impossible), the prior boundaries are ten standard deviations away from the posterior median such that the effect is negligible.
If instead we truncate the posterior on $[a,b]$, the expression for the Bayes factor that generalizes Eq.~(\ref{eq: Bayes factor simple}) has an extra factor
$[\erf(\frac{b-\nu}{\sqrt{2}\tau}) - \erf(\frac{a-\nu}{\sqrt{2}\tau})] / [\erf(\frac{b-\hat{x}}{\sqrt{2}\hat{\sigma}}) - \erf(\frac{a-\hat{x}}{\sqrt{2}\hat{\sigma}})]$,
where
$\nu = (\mu\hat{\sigma}^2 + \hat{x}\sigma^2) / (\hat{\sigma}^2 + \sigma^2)$,
and
$\tau = \hat{\sigma}\sigma / \sqrt{\hat{\sigma}^2 + \sigma^2}$.
If the Gaussian priors were also truncated on $[a,b]$ then we would have $\mathcal{B}_{A/U}\to1$ as $\sigma\to\infty$.

\subsection{Comparison to heuristic fraction}

\begin{figure}
\includegraphics[width=\columnwidth]{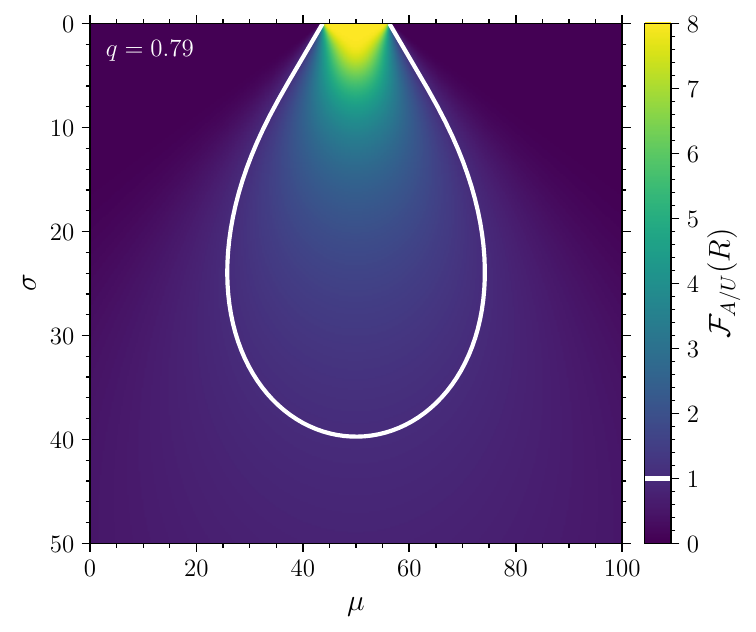}
\caption{
Relative heuristic fractions $\mathcal{F}_{A/U}(R)$ in a simple example for a univariate parameter $x$, to be compared against the Bayes factor $\mathcal{B}_{A/U}$ in Fig.~\ref{fig: simple}.
We assume a Gaussian posterior with mean $\hat{x}=50$ and standard deviation $\hat{\sigma}=5$, and uniform prior between $a=0$ and $b=100$.
We consider alternative Gaussian priors with means $\mu$ and standard deviations $\sigma$ and mark in white the equal preference $\mathcal{F}_{A/U}(R)=1$ boundary inside which the Gaussian prior is favoured.
The region $R$ is chosen to be the $q=79\%$ symmetric CI of the posterior.
}
\label{fig: relative}
\end{figure}

In Fig.~\ref{fig: relative} we show the relative heuristic fraction $\mathcal{F}_{A/U}(R)$ from Eq.~(\ref{eq: relative fraction}) for the same univariate example.
We choose the region $R$ to be the symmetric 79\% ($q=0.79$) credible interval of the posterior; this value was chosen as it results in $\max\mathcal{F}_{A/U}(R) \approx \max\mathcal{B}_{A/U}$ for this specific example.
The heuristic fraction can be expressed entirely in terms of error functions;
from Eq.~(\ref{eq: heuristic fraction}) we have that the fraction of prior samples from $\pi(x|A)$ inside the posterior region $R$ is
\begin{align}
f_A(R)
=
\frac{1}{2} \bigg[ &\erf \left( \frac{\hat{x} - \mu + \sqrt{2}\hat{\sigma}\erf^{-1}(q)}{\sqrt{2}\sigma} \right)
\nonumber \\
+ &\erf \left( \frac{\mu - \hat{x} + \sqrt{2}\hat{\sigma}\erf^{-1}(q)}{\sqrt{2}\sigma} \right) \bigg]
\, .
\label{eq: heuristic fraction simple}
\end{align}
A similar calculation for the flat prior $\pi(x|U)$ therefore yields for the relative fraction in Eq.~(\ref{eq: relative fraction})
\begin{align}
\mathcal{F}_{A/U}(R)
=
\frac{ (b-a)f_A(R) } { 2\sqrt{2}\hat{\sigma}\erf^{-1}(q) }
\, .
\label{eq: relative fraction simple}
\end{align}

The relative fraction saturates much earlier for narrow models, i.e., $\sigma\approx0$, compared to the Bayes factor, for which there is a unique maximum.
There is a sharp transition from $\mathcal{F}_{A/U}(R) = 0$ to $\mathcal{F}_{A/U}(R) = \max\mathcal{F}_{A/U}(R)$ at $\sigma=0$ because the gradient of the posterior density is neglected in the heuristic computation; the model lies entirely within the posterior region $R$ once $\mu$ is within the symmetric interval defined by $q$.
The contours of constant $\mathcal{F}_{A/U}(R)$ also do not capture the full behaviour of the $\mathcal{B}_{A/U}$ contours, in particular for small $\sigma$.
Otherwise, $\mathcal{F}_{A/U}(R)$ represents a reasonable approximation of $\mathcal{B}_{A/U}$ in this simple example but, most importantly, the quantile $q$ was hand picked to result in relative fractions similar to the Bayes factor.
If one were to select a different quantile, e.g., the arbitrary choice of 90\%, the resulting distribution of relative heuristic fractions in Fig.~\ref{fig: relative} would be altered and, in particular, less similar to the Bayes factors in Fig.~\ref{fig: simple}.

\begin{figure}
\includegraphics[width=\columnwidth]{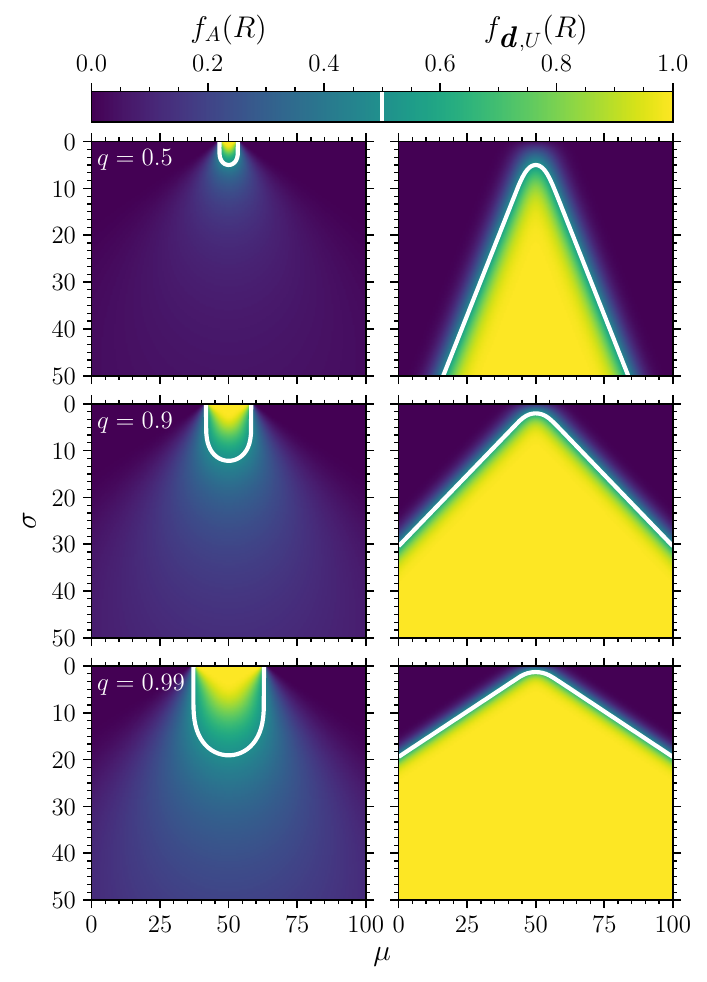}
\caption{
Heuristic fractions in the simple univariate Gaussian example, to be compared against the Bayes factor $\mathcal{B}_{A/U}$ in Fig.~\ref{fig: simple}.
The left (right) column shows the fraction $f_A(R)$ ($f_{\vec{d},U}(R)$) of model (posterior) samples inside the $q$ symmetric CI of the posterior (model).
The quantile $q=50\%,90\%,99\%$ is increased down the rows.
White lines indicate the arbitrarily chosen $f_A(R), f_{\vec{d},U}(R) = 1/2$ boundaries.
}
\label{fig: heuristic}
\end{figure}

In Fig.~\ref{fig: heuristic} we show the values of the heuristic confidence fractions $f_A(R)$ -- which counts the fraction of model $A$ samples inside a region $R$ of the posterior -- and $f_{\vec{d},U}(R)$ -- which counts the fraction of posterior samples inside a chosen region of the model prior.
We again take the regions $R$ as symmetric CIs.
We display the fractions as functions of both $\mu$ and $\sigma$ and $R$ by selecting the $q=50\%,90\%, 99\%$ symmetric intervals.
The expression for $f_{\vec{d},U}(R)$ is found by interchange of $\hat{x}$ and $\hat{\sigma}$ with $\mu$, and $\sigma$, respectively, in Eq.~(\ref{eq: heuristic fraction simple}).
Figure~\ref{fig: heuristic} shows that these two fractions are indeed different.
While $f_A(R)$ is maximized for narrow models ($\sigma\to0$), $f_{\vec{d},U}(R)$ is maximized for broad models ($\sigma\to\infty$).
In the former case, narrow alternative priors $\pi(x|A)$ place more (less) density inside $R$ if the mean $\mu$ is inside (outside) this region.
On the other hand, in this case broadening $\pi(x|A)$ places more density outside $R$ and thus $f_A(R)\to0$ as $\sigma\to\infty$.
In the latter case, broadening the prior encapsulates more of the posterior support and thus $f_{\vec{d},U}(R)\to1$ as $\sigma\to\infty$.
The two disparate fractions agree only when there is no overlap between the alternative prior and the posterior, as seen in the top left and right corners of the panels in Fig.~\ref{fig: heuristic}.
The heuristic fraction $f_A(R)$ represents the Bayes factor much worse than the relative fraction $\mathcal{F}_{A/U}(R)$, in particular in the contours of constant $f_A(R)$.
This also highlights the dependence on the arbitrary choice of $R$ through $q$, which changes the variation of the heuristic fractions over the parameters $\mu$ and $\sigma$.
As $q\to1$, we find $f_A(R), f_{\vec{d},U}(R) \to 1$ since in the present example the posterior and priors have unbounded supports.

\end{document}